%% file: distill.tex
\documentclass{article}
\usepackage{ijcai19}

\usepackage{times}
\usepackage{soul}
\usepackage{url}
\usepackage[hidelinks]{hyperref}
\usepackage[utf8]{inputenc}
\usepackage[small]{caption}
\usepackage{graphicx}
\usepackage{amsmath}
\usepackage{booktabs}
\usepackage{arydshln}
\usepackage{amsfonts}
\usepackage{microtype}
\usepackage{nicefrac}

\usepackage{subcaption}

\usepackage{multirow}
\usepackage{tablefootnote}
\usepackage[hang,flushmargin]{footmisc}   
\usepackage[linesnumbered,ruled,vlined]{algorithm2e}
\usepackage{cleveref}

\title{The Pupil Has Become the Master: Teacher-Student Model-Based\\Word Embedding Distillation with Ensemble Learning}

\author{
Bonggun Shin$^1$
\and
Hao Yang$^2$\And
Jinho D. Choi$^{1}$
\affiliations
$^1$Department of Computer Science, Emory University, Atlanta, GA\\
$^2$Visa Research, Palo Alto, CA\\
\emails
\texttt{bonggun.shin@emory.edu},
\texttt{haoyang@visa.com},
\texttt{jinho.choi@emory.edu}
}

\begin{document}

\title{The Pupil Has Become the Master: Teacher-Student Model-Based\\Word Embedding Distillation with Ensemble Learning}

\maketitle

\input{tex/abstract}
\input{tex/introduction}
\input{tex/background}
\input{tex/distillation}

\input{tex/ensemble}

\input{tex/experiments}

\input{tex/conclusion}

\bibliography{distill}
\bibliographystyle{named}

\appendix

\end{document}

%% file: tex/abstract.tex
\begin{abstract}
Recent advances in deep learning have facilitated the demand of neural models for real applications.
In practice, these applications often need to be deployed with limited resources while keeping high accuracy.
This paper touches the core of neural models in NLP, word embeddings, and presents a new embedding distillation framework that remarkably reduces the dimension of word embeddings without compromising accuracy.
A novel distillation ensemble approach is also proposed that trains a high-efficient student model using multiple teacher models.
In our approach, the teacher models play roles only during training such that the student model operates on its own without getting supports from the teacher models during decoding, which makes it eighty times faster and lighter than other typical ensemble methods.
All models are evaluated on seven document classification datasets and show significant advantage over the teacher models for most cases.
Our analysis depicts insightful transformation of word embeddings from distillation and suggests a future direction to ensemble approaches using neural models.
\end{abstract}

%% file: tex/introduction.tex
\section{Introduction}
As deep learning starts dominating the field of machine learning, there have been growing interests in deploying deep neural models for real applications.
\cite{hinton2015distilling} stated that academic research on model development had mostly focused on accuracy improvement, whereas the deployment of deep neural models would also require the optimization of other practical aspects such as speed, memory, storage, power, etc.
To satisfy these requirements, several neural model compression methods have been proposed, which can be categorized into the following four:
weight pruning~\cite{denil2013predicting,han2015learning,jurgovsky2016evaluating}, 
weight quantization~\cite{han2015deep,jurgovsky2016evaluating,ling2016word}, 
lossless compression~\cite{van1976construction,han2015learning}, 
and distillation~\cite{mou2016distilling}.
This paper focuses on distillation methods that can remarkably reduce the model size, resulting in much less memory usage and fewer computations.

\noindent Distillation aims to extract core elements from a complex network and transfer them to a simpler network so it gives comparable results to the complex network.
It has been shown that the core elements can be transferred to various types of networks i.e., deep to shallow networks~\cite{ba2014deep}, recurrent to dense networks~\cite{chan2015transferring}, and vice versa~\cite{romero2014fitnets,tang2016recurrent}. 
Lately, embedding distillation was suggested~\cite{mou2016distilling}, which transferred the output of the projection layer in the source network as input to the target network, although accuracy drop was expected with this approach.
Considering the upper bound of a distilled network, that is the accuracy achieved by the original network~\cite{ba2014deep}, enough room is left for the improvement of embedding distillation. 
Distilled embeddings can significantly enhance the efficiency of deep neural models in NLP, where the majority of model space is occupied by word embeddings.
\begin{figure*}[hbtp!]
	\centering
	\begin{subfigure}[b]{0.3\textwidth}
		\centering
		\includegraphics[scale=0.059]{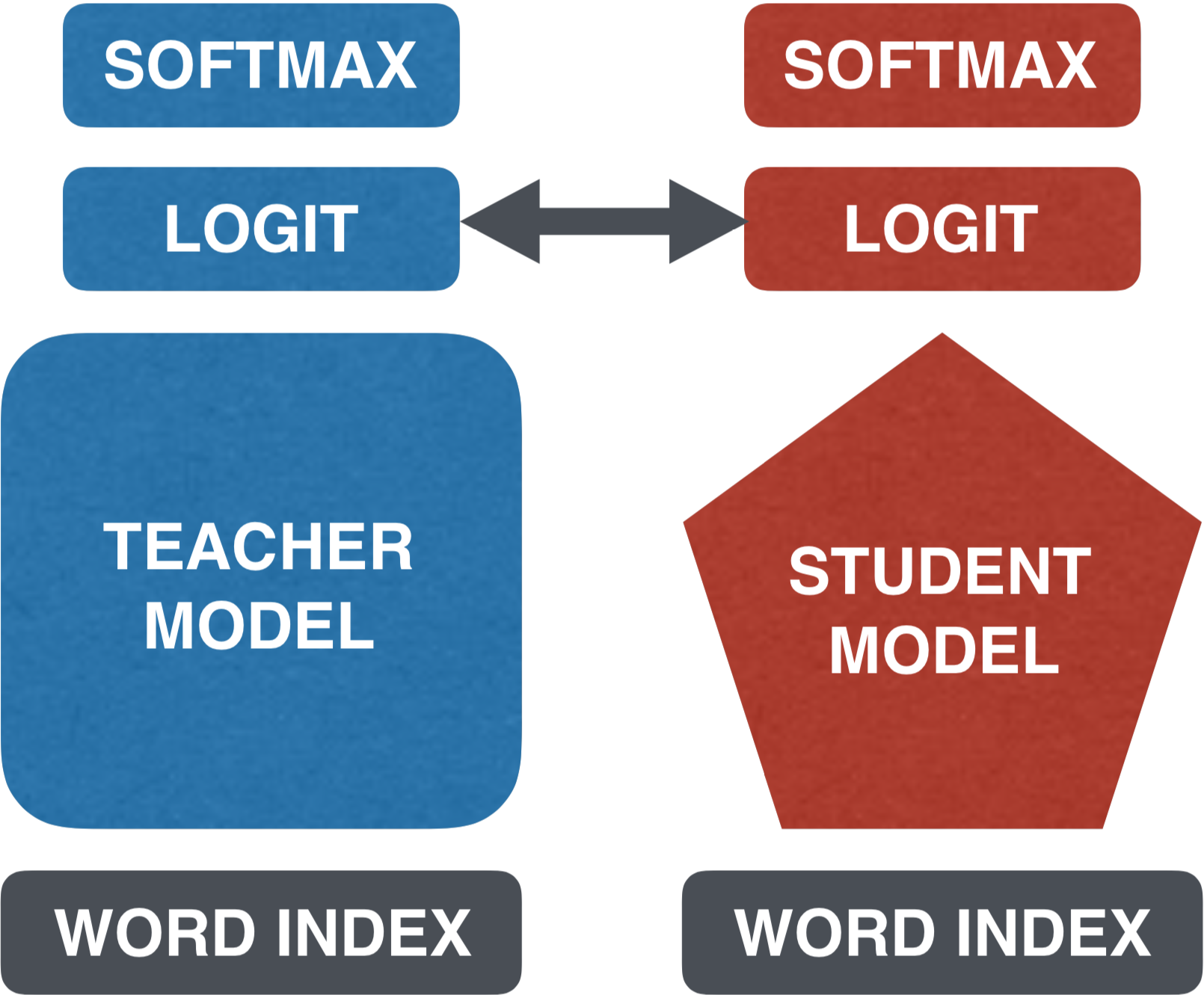}
		\caption{Logit Matching (Sec.~\ref{ssec:logit})}
		\label{fig:model.lm}
	\end{subfigure}%
	~~~~~~~~ 
	\begin{subfigure}[b]{0.3\textwidth}
		\centering
		\includegraphics[scale=0.065]{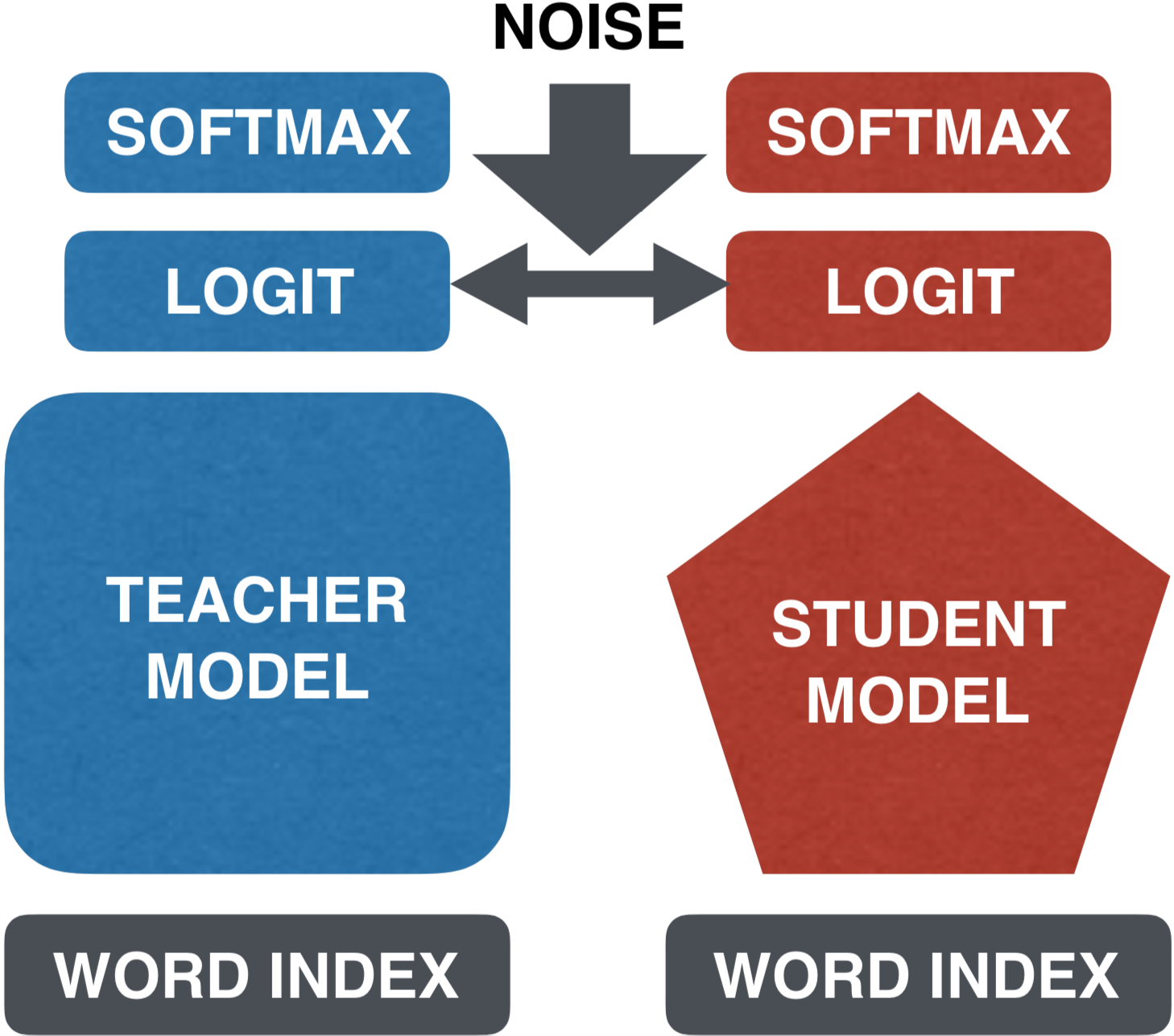}
		\caption{Noisy Logit Matching (Sec.~\ref{ssec:nlogit})}
		\label{fig:model.nlm}
	\end{subfigure}%
	~~~~~~~~
	\begin{subfigure}[b]{0.3\textwidth}
		\centering
		\includegraphics[scale=0.065]{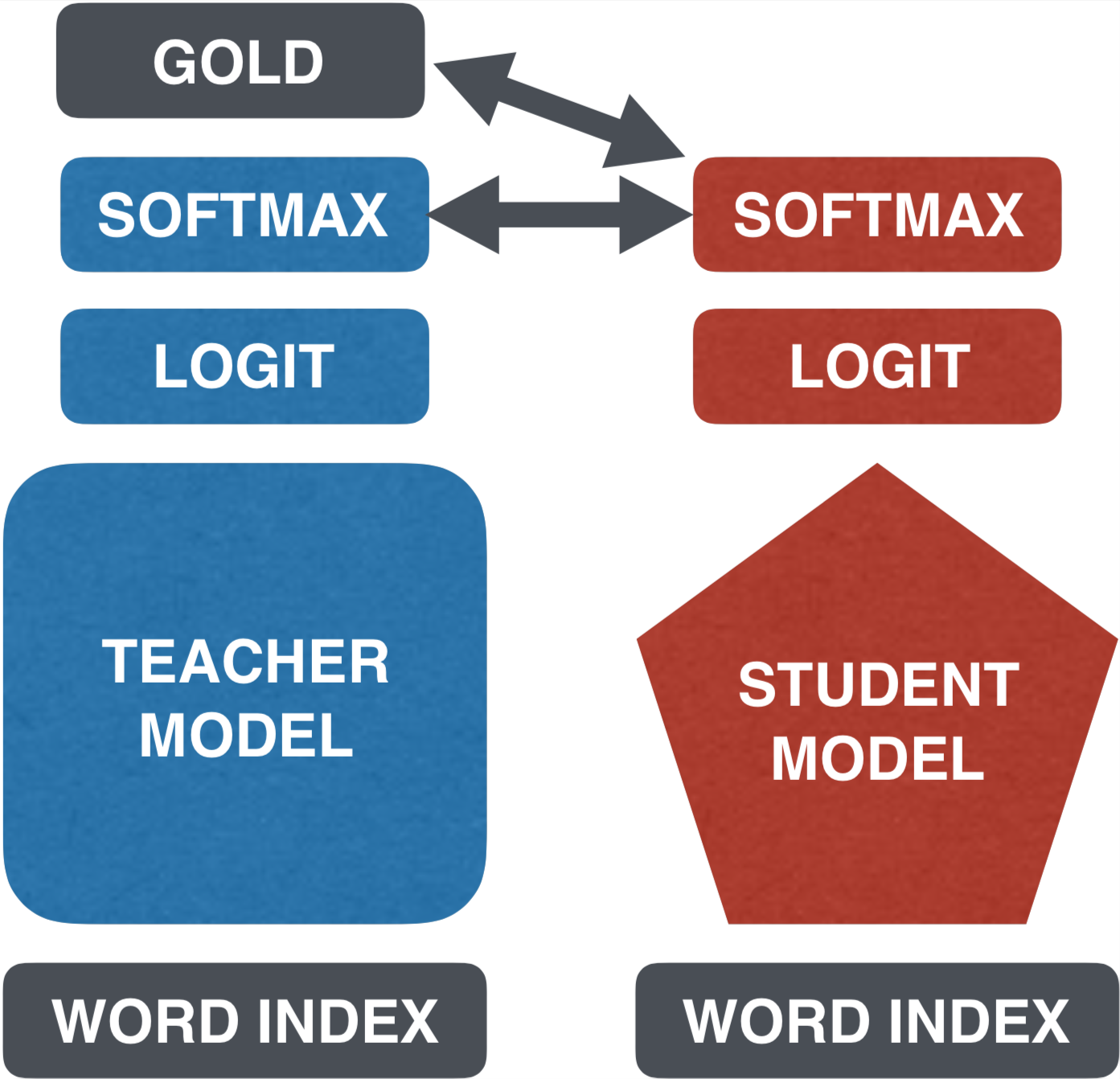}
		\caption{Softmax Tau Matching (Sec.~\ref{ssec:stm})}
		\label{fig:model.tsm}
	\end{subfigure}
	\caption{Three teacher-student methods described in Background section, which uses different cost functions to transfer trained knowledge from the teacher model to the student model.}
	\label{fig:teachter-student}
	\vspace{-1ex}
\end{figure*}

In this paper, we first propose a new embedding distillation method based on three teacher-student frameworks, which is a more advanced way of embedding distillation, because the previous one~\cite{mou2016distilling} is a standalone embedding distillation (Section~\ref{ssec:enc}) with limited knowledge transfer. Our distilled embeddings not only enable the target network to outperform the previous state of the art~\cite{mou2016distilling}, but also are eight times smaller than the original word embeddings yet allow the target network to achieve compatible (sometimes higher) accuracy to the source network.
We then present a novel ensemble approach which extends this distillation framework by allowing multiple teacher models when training a student model. After learning from multiple teachers during training, the student model runs on its own during decoding such that it performs faster and lighter than any of the teacher models yet pushes the accuracy much beyond them with just 1.25\% ($\nicefrac{50}{4000}$) the size of other typical ensemble models. All models are evaluated on seven document classification datasets; our experiments show the effectiveness of the proposed frameworks, and our analysis illustrates an interesting nature of the distilled word embeddings.
To the best of our knowledge, this is the first time that embedding distillation is thoroughly examined for natural language processing and used in ensemble to achieve such promising results.\footnote{Our code is publicly available\\: \url{https://github.com/bgshin/distill_demo}}




%% file: tex/background.tex
\section{Background}
\label{sec:background}
\vspace{-0.5ex}
Our embedding distillation framework is based on teacher-student models \cite{ba2014deep,sau2016deep,hinton2015distilling}, where teacher models are trained on deep neural networks and transfer their knowledge to student models on simpler networks.
The following subsections describe three popular teacher-student methods applied to our framework.
The main difference between these three methods is in their cost functions (Figure~\ref{fig:teachter-student}).
The last subsection discusses embedding encoding that is used to extract distilled embeddings from the projection layer.

Throughout this section, a \textit{logit} refers to a vector representing the layer immediately before the softmax layer in a neural network, where $z_i$ and  $v_i$ are the teacher's and student's logit values for the class $i$, respectively.
Note that the student models are not necessarily optimized for only the gold labels but also optimized for the logit values from the teacher models in these methods.

\subsection{Logit Matching (LM)}
\label{ssec:logit}

Proposed by \cite{ba2014deep}, the cost function of this teacher-student method is defined by the logit differences between the teacher and the student models ($D$: the total number of classes):\vspace{-2.5ex}
$$L_{LM} = \frac{1}{2\cdot D}\sum_{i=1}^D|z_i-v_i |^2$$

\subsection{Noisy Logit Matching (NLM)}
\label{ssec:nlogit}

Proposed by \cite{sau2016deep}, this method is similar to Logit Matching except that Gaussian noise is introduced during the distillation, simulating variations in the teacher models, 
which gives a similar effect for the student to learn from multiple teachers.
The cost function takes random noise $\eta$ drawn from Gaussian distribution such that the logit of each teacher model is $z_{i}' = (1+\eta)\cdot z_{i}$.
Thus, the final cost function becomes $L_{NLM} = \frac{1}{2D}\sum_{\forall i}|z_i'-v_i |^2$.

\subsection{Softmax Tau Matching (STM)}
\label{ssec:stm}

Proposed by \cite{hinton2015distilling}, this method is based on softmax matching where softmax values are compared between the teacher and the student models instead of logits.
Later, \cite{hinton2015distilling} added two hyperparameters to further generalize this method.
The first hyperparameter, $\lambda$, is for the weighted average of two sub-cost functions, where the first sub-cost function measures a cross-entropy between the student's softmax predictions and the truth values, represented as $L_{1} = -\sum_{\forall i}  y_{i}\log{p_{i}}$ ($i$  indexes classes, $y$ is the gold label, $p_{i} \in (0,1)$ is the prediction for a sample).
Another cost function involves the second hyperparameter, $\tau$, that is a temperature variable normalizing the output of the teacher's logit value:
\vspace{-1ex}
$$s_i(z,\tau) = \frac{ e^{\nicefrac{z_i}{\tau}} }{\sum_{j=1}^D e^{\nicefrac{z_j}{\tau}}}$$
\noindent Given $s_i$, the second cost function can be defined as $L_{2} = -\sum_{\forall i}  s_{i}(z,\tau) \log{p_{i}}$.
Therefore, the final cost function becomes $L_{STM} = \lambda L_{1} + (1-\lambda) L_{2}$.
If $\lambda$ weights more on $L_1$, the student model values more on the gold labels than teacher's predictions.
If $\tau$ is greater, the teacher's output becomes more uniformed, implying that the probability values are spread out more throughout all classes.

\subsection{Embedding Encoding (ENC)}
\label{ssec:enc}

\begin{figure}[ht!]
	\centering
	\begin{subfigure}[b]{0.23\textwidth}
		\centering
		\includegraphics[scale=0.065]{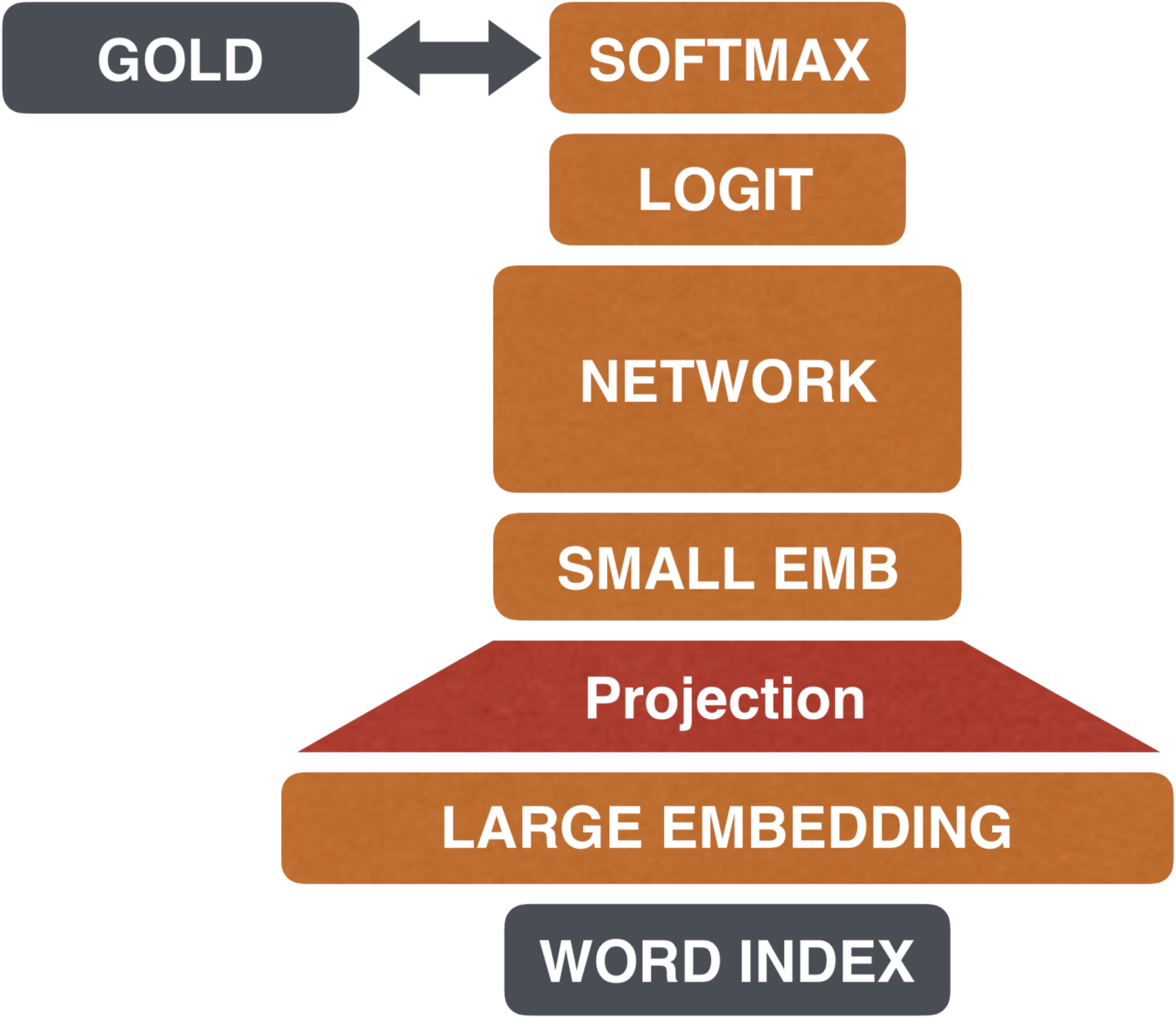}
		\caption{Embedding distillation.}
		\label{fig:embedding-encoding-distill}
	\end{subfigure}%
	~ 
	\begin{subfigure}[b]{0.23\textwidth}
		\centering
		\includegraphics[scale=0.066]{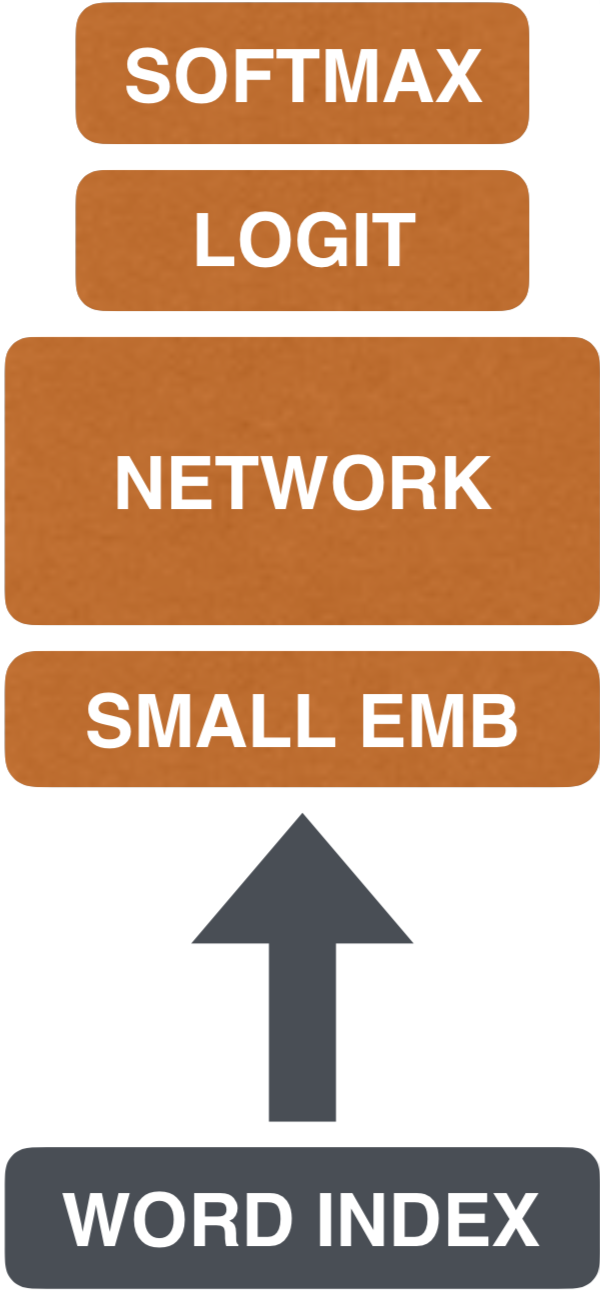}
		\caption{Model deployment.}
		\label{fig:embedding-encoding-deploy}
	\end{subfigure}%
	\caption{(a) The core elements from large embeddings are distilled during training using gold labels and transferred to smaller embeddings. (b) Only the small embeddings are kept for the deployment, resulting less space and fewer computations.}
	\label{fig:embedding-encoding}
\end{figure}

Embedding distillation was first proposed by \cite{mou2016distilling} for NLP tasks.
Unlike our framework, their method does not rely on teacher-student models, but rather directly trains a single model with an encoding layer inserted between the embedding layer and its upper layer in the network (Figure~\ref{fig:embedding-encoding-distill}).
Each word $w_i$ is entered to an embedding layer $\phi$ that yields a large embedding vector $\phi(w_i)$.
This vector is projected into a smaller embedding space by $W_{enc}$ with an activation function $f$.
As a result, a smaller embedding vector $\phi'(w_i)$ is produced for $w_i$ as follows ($b_{enc}$: a bias for the projection):
$$\phi'(w_i) = f(W_{enc} \cdot \phi (w_i)+ b_{enc})$$
The smaller embedding $\phi'(w_i)$ generated by this projection contains distilled knowledge from the larger embedding $\phi(w_i)$.
The cost function of this method simply measures the cross-entropy between gold labels and
the softmax output values.

%% file: tex/distillation.tex

\section{Embedding Distillation}
\label{sec:distillation}

\begin{figure}[htbp!]
    \centering
   	\begin{subfigure}[b]{0.23\textwidth}
   		\centering
		\includegraphics[scale=0.065]{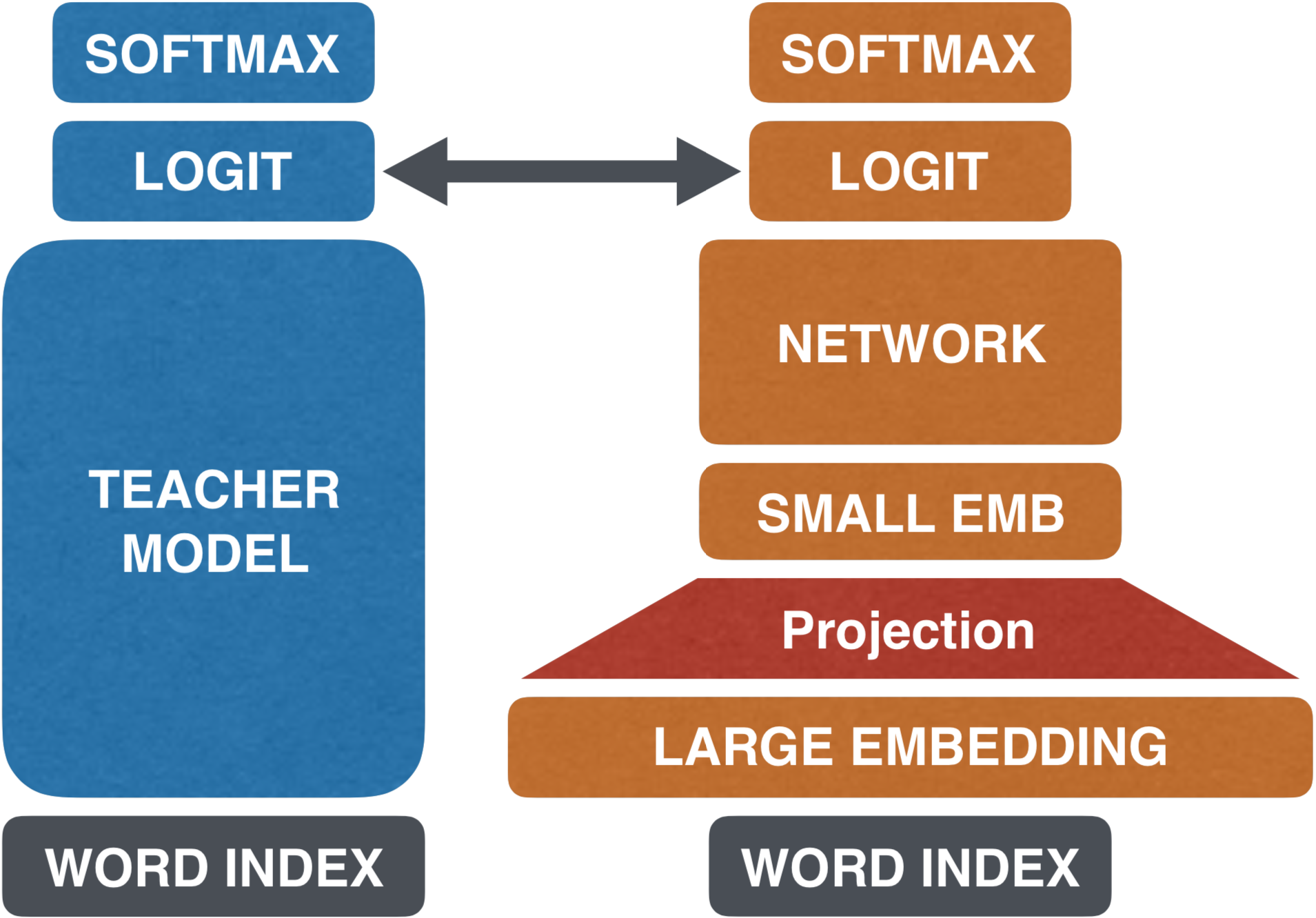}
   		\caption{Embedding distillation.}
   		\label{fig:embedding-distill}
   	\end{subfigure}%
   	~ 
   	\begin{subfigure}[b]{0.23\textwidth}
   		\centering
   		\includegraphics[scale=0.066]{images/ENCTST.png}
   		\caption{Model deployment.}
   		\label{fig:embedding-deploy}
   	\end{subfigure}%
   	\caption{(a) Our proposed embedding distillation framework using teacher-student models.
             (b) Only the small embeddings are kept for deployment, similarly to Figure~\ref{fig:embedding-encoding-deploy}.}
         	\vspace{-2ex}
	\label{fig:lm}
\end{figure}

Our proposed embedding distillation framework begins by training a teacher model using the original embeddings.
After training, the teacher model generates the corresponding logit value for each input in the training data.
Then, a student model that comprises a projection layer is optimized for the logit (or softmax) values from the teacher model.
After training the student model, small embeddings are distilled from the projection layer (Figure~\ref{fig:embedding-distill}).

\noindent The original large embeddings as well as weights in the projection layer are discarded for deployment such that the small embeddings can be referenced directly from the word indices in the student model during decoding (Figure~\ref{fig:embedding-deploy}).
Such distillation significantly reduces the model size and computations in the network, which is welcomed in production.



\subsection{Distillation via Teacher-Student Models}
\label{ssec:distillation}

\noindent Projecting a vector into a lower dimensional space generally entails information loss, although it does not have to be the case under two conditions.
First, the source embeddings comprise both relevant and irrelevant information for the target task, therefore, there is a room to discard the irrelevant information.
Second, the projection layer in the target network is capable of preserving the relevant information from the source embeddings.
The first condition is met for NLP because most vector space models such as Word2Vec~\cite{mikolov2013distributed}, Glove~\cite{pennington-socher-manning:2014:EMNLP2014}, or FastText~\cite{Q17-1010} are trained on a vast amount of text, where only a small portion is germane to a specific task.

To meet the second condition, teacher-student models are adapted to our distillation framework, where the projection layer in a student model learns the relevant information for the target task from a teacher model.
The output dimension of the projection layer is generally much smaller than the size of the embedding layer in a teacher model.
Note that it is possible to integrate multiple projection layers in a student model; Experiment section discusses performance difference by adding different numbers of projection layers in the student model.

\subsection{Projection Layer Initialization}
\label{ssec:encoding}

Unlike \cite{mou2016distilling} who randomly initialized the projection layer, it is initialized with vectors pre-trained by an autoencoder~\cite{hinton1994autoencoders} in our framework.
This initialization stabilizes and improves optimization of neural networks during training, resulting more robust models.



%% file: tex/ensemble.tex
\section{Distillation Ensemble}
\label{sec:ensemble}

Ensemble methods generally achieve higher accuracy than a standalone model; however, slow speed is expected due to the runs from multiple models in ensemble.
This section presents a novel ensemble approach based on our distillation framework using logit matching that produces a light-weighted student model trained by multiple teachers (Figure~\ref{fig:distill-ensemble}).

The premise of this approach is that it is possible to have multiple teacher models train a student model by combining their logit values such that the student no longer needs the teachers during decoding because it already learned ``enough'' from them during training.
As a result, our ensemble approach ensures higher efficiency for the student model than for any of the teacher models during decoding.
The following sections describe two different ensemble methods applied to our framework.

	\vspace{-1ex}
\begin{figure}[htbp!]
	\centering
	\begin{subfigure}[b]{0.25\textwidth}
		\centering
		\includegraphics[scale=0.082]{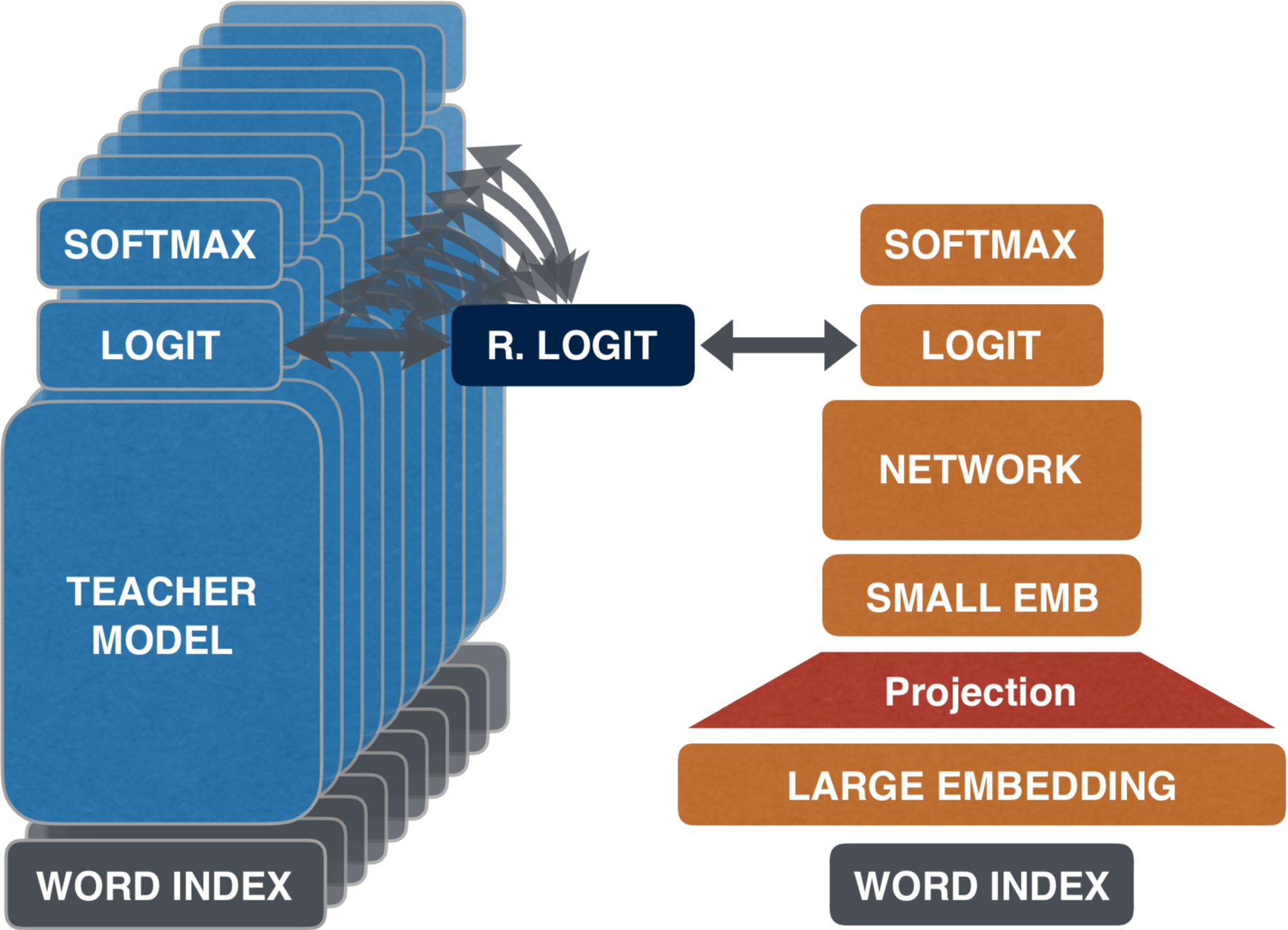}
		\caption{Embedding distill ensemble.}
		\label{fig:distill-ensemble}
	\end{subfigure}%
	~ 
	\begin{subfigure}[b]{0.23\textwidth}
		\centering
		\includegraphics[scale=0.065]{images/ENCTST.png}
		\caption{Model deployment.}
		\label{fig:distill-ensemble-deploy}
	\end{subfigure}%
	\caption{(a) Our proposed embedding distillation ensemble model. One representing logit (R. LOGIT) is calculated from a set of multiple teachers' logits by the proposed ensemble methods.
		(b) No need to evaluate teachers at deployment, unlike other ensemble methods.}
	\label{fig:distill-ensemble}
	\vspace{-4ex}
\end{figure}



\subsection{Routing by Agreement Ensemble (RAE)}
\label{sssec:rae-ensemble}

This method gives more weights to the majority, by adopting the dynamic routing algorithm presented by \cite{sabour2017dynamic}.
It first collects the consensus of all teachers, then boosts weights of the teachers who strongly agree with that consensus
whereas suppresses the influence of the teachers who do not agree as much.
The procedure of calculating the representing logit is described in Algorithm~\ref{alg:rae}.

The squash function in lines 4 and 9 is a non-linear activation function that ensures the norm of the output vector to be in $[0,1]$~\cite{sabour2017dynamic}.
This vectorized activation is important for a routing algorithm because both magnitude and direction play an important role in the agreement calculation by the dot product, which enforces the consensus into the direction with strong confidence.


\subsection{\fontsize{10.5pt}{12.6pt}\selectfont Routing by Disagreement Ensemble (RDE)}
\label{sssec:rde-ensemble}

This method focuses on the minority vote instead, because minority opinions may cast important information for the task.
The algorithm is the same as RAE, except for the sign of the weight update (line 2 in Algorithm~\ref{alg:rae}).

\begin{algorithm}[hbtp!]{
	\small
	\SetAlgoLined
    \SetAlgoVlined
    \DontPrintSemicolon
	\KwIn{Teachers' logits $Z \in \mathbb{R}^{T \times C}$, and\linebreak
		an algorithm selector $b \in \{$RAE, RDE$\}$}.
	\KwOut{The representing logit, $z^{rep} \in \mathbb{R}^{C}$.}
	$k\leftarrow1$ \textbf{if} {$b$ is RAE} \textbf{else} {$-1$}\;
	\For{$t \in \{1, \ldots, T\}$}
	{
		$x^t \leftarrow $ squash$(z^t)$, 		$w_t \leftarrow 0$\;
	}
	\While{$n$ iterations}{
		$c \leftarrow $ softmax$(w)$\;
		$z^{rep} = \sum_{t=1}^{T} c_{t} \cdot z^{t}$\;
		$s \leftarrow $ squash$(z^{rep})$\;
		\If{not last iteration}
		{
			\For{$t \in \{1, \ldots, T\}$}
			{
				$w_t \leftarrow w_t + k x^t \cdot s $\;
			}
		}
		
	}
	\Return $z^{rep}$
	\caption{\small Get R. LOGIT for RAE and RDE}
	\label{alg:rae}
}
\end{algorithm}


%% file: tex/experiments.tex
\section{Experiments}
\label{sec:experiment}
\subsection{Datasets}
All models are evaluated on seven document classification datasets in Table~\ref{tbl:data}.
MR, SST-*, and CR are targeted at the task of sentiment analysis while Subj, TREC, and MPQA are targeted at the classifications of subjectivity, question types, and opinion polarity, respectively.
About 10\% of the training sets are split into development sets for SST-* and TREC, and about 10/20\% of the provided resources are divided into development/evaluation sets for the other datasets, respectively.

\begin{table}[htbp!]
	\centering\resizebox{\columnwidth}{!}{
	\begin{tabular}{l||r|r|r|r}
		\multicolumn{1}{c||}{\bf Dataset} & \multicolumn{1}{c|}{\bf C} & \multicolumn{1}{c|}{\textbf{\texttt{TRN}}} & \multicolumn{1}{c|}{\textbf{\texttt{DEV}}}& \multicolumn{1}{c}{\textbf{\texttt{TST}}}\\  \hline \hline
		MR~\cite{pang2005seeing}         & 2 & 7,684 &   990 & 1,988 \\
		SST-1~\cite{socher2013recursive} & 5 & 8,544 & 1,101 & 2,210 \\
		SST-2~\cite{socher2013recursive} & 2 & 6,920 &   872 & 1,821 \\
		Subj~\cite{pang2004sentimental}  & 2 & 7,199 &   907 & 1,894 \\
		TREC~\cite{li2002learning}       & 6 & 4,952 &   500 &   500  \\
		CR~\cite{hu2004mining}           & 2 & 2,718 &   340 &   717  \\
		MPQA~\cite{wiebe2005annotating}  & 2 & 7,636 &   955 & 2,015
	\end{tabular}}
	\caption{Seven datasets used for our experiments. C: number of classes, \texttt{TRN}/\texttt{DEV}/\texttt{TST}: number of instances in training/development/evaluation set.}
	\label{tbl:data}
\vspace{-2ex}
\end{table}

\subsection{Word Embeddings}
For sentiment analysis, raw text from the Amazon Review dataset\footnote{\texttt{http://snap.stanford.edu/data/web-Amazon.html}} is used to train word embeddings, resulting 2.67M word vectors.
For the other tasks, combined text from Wikipedia and the New York Times Annotated corpus\footnote{\texttt{https://catalog.ldc.upenn.edu/LDC2008T19}} are used, resulting 1.96M word vectors.
For each group, two sets of embeddings are trained with dimensions of 50 and 400 by Word2Vec~\cite{mikolov2013distributed}.
While training, default hyper-parameters are used without an explicit hyper-parameter tuning.


\begin{table*}[htb!]
	\centering\resizebox{\textwidth}{!}{
		\begin{tabular}{l:c||c|c|c|c|c|c|c}
			\bf Model     & \bf Emb.Size &         \bf MR          &        \bf SST-1         &        \bf SST-2         &         \bf Subj         &         \bf TREC         &          \bf CR          & \multicolumn{1}{c}{\bf MPQA} \\ \hline\hline
			ENC           &  50  &   77.11$\pm$ 0.97      &     44.94$\pm$ 1.26      &     83.71$\pm$ 1.41      &     90.64$\pm$ 0.49      &     90.60$\pm$ 1.10      &     80.88$\pm$ 1.22      &       88.65$\pm$ 0.60        \\ \hline\hline
			CNN-400       &   400  &          79.07           &          49.86           &          86.22           &          92.34           &          93.60           &          83.82           &            88.78             \\
			CNN-50        &     50  &       78.07           &          45.07           &          84.51           &          90.81           &          91.00           &          80.89           &            86.40             \\ \hline
			LM+TSED        &    50  &   77.63$\pm$ 0.37      &     48.71$\pm$ 0.73      &     85.04$\pm$ 0.64      &     91.91$\pm$ 0.29      &     92.48$\pm$ 0.73      &     81.84$\pm$ 0.57      &   89.14$\pm$ 0.25   \\
			NLM+TSED       &   50  & 78.10$\pm$ 0.40 &     48.66$\pm$ 0.83      & 85.17$\pm$ 0.16 &     92.03$\pm$ 0.36      &     92.36$\pm$ 0.91      & 83.04$\pm$ 0.74 &       88.90$\pm$ 0.34        \\
			STM+TSED       &     50  &  77.81$\pm$ 0.33      & 49.10$\pm$ 0.34 &     84.72$\pm$ 0.70      & 92.25$\pm$ 0.38 & 92.76$\pm$ 0.65 &     80.31$\pm$ 0.42      &       89.13$\pm$ 0.28        \\ \hline
			LM+TSED+PT+2L  &   50  &\textbf{79.06$\pm$ 0.59} &     49.82$\pm$ 0.54      &     85.75$\pm$ 0.42      & \textbf{92.63$\pm$ 0.23} &     92.58$\pm$ 0.86      &     83.40$\pm$ 0.76      &       89.44$\pm$ 0.20        \\
			NLM+TSED+PT+2L &    50  &   78.60$\pm$ 0.70      & \textbf{49.90$\pm$ 0.59} &     85.31$\pm$ 0.75      &     92.26$\pm$ 0.20      &     92.80$\pm$ 0.62      & \textbf{83.82$\pm$ 0.49} &       89.61$\pm$ 0.14        \\
			STM+TSED+PT+2L &   50  &    78.77$\pm$ 0.70      &     49.19$\pm$ 0.71      & \textbf{85.83$\pm$ 0.59} &     92.38$\pm$ 0.53      & \textbf{93.48$\pm$ 0.30} &     83.57$\pm$ 0.85      &   \textbf{89.95$\pm$ 0.28}   \\ \hline\hline
			LSTM-400      &    400  &        79.28           &          49.23           &          86.22           &          92.71           &          92.00           &          82.98           &            89.73             \\
			LSTM-50       &     50  &       77.16           &          43.76           &          83.36           &          90.02           &          86.00           &          80.06           &            85.66             \\ \hline
			LM+TSED        &  50  &     78.61$\pm$ 0.80      &     48.79$\pm$ 0.27      &     85.81$\pm$ 0.77      &     91.74$\pm$ 0.36      & 92.56$\pm$ 0.89 &     82.76$\pm$ 0.36      &       89.59$\pm$ 0.28        \\
			NLM+TSED       &  50  & 78.89$\pm$ 0.73 & 48.81$\pm$ 0.44 &     85.55$\pm$ 0.74      &     91.87$\pm$ 0.32      &     91.80$\pm$ 1.29      & 82.96$\pm$ 0.50 &   89.63$\pm$ 0.10   \\
			STM+TSED       &  50  &     78.85$\pm$ 0.60      &     48.77$\pm$ 0.83      & 86.11$\pm$ 0.36 & 91.99$\pm$ 0.16 &     92.36$\pm$ 0.43      &     82.96$\pm$ 0.72      &       89.60$\pm$ 0.16        \\ \hline
			LM+TSED+PT+2L  &  50  & \textbf{80.33$\pm$ 0.40} & \textbf{49.37$\pm$ 0.39} &     86.12$\pm$ 0.47      & \textbf{92.53$\pm$ 0.29} &     91.91$\pm$ 0.63      & \textbf{83.54$\pm$ 0.78} &   \textbf{90.15$\pm$ 0.13}   \\
			NLM+TSED+PT+2L &    50  &   79.33$\pm$ 0.66      &     48.87$\pm$ 0.53      &     85.89$\pm$ 0.43      &     92.35$\pm$ 0.18      &     92.08$\pm$ 0.46      &     83.32$\pm$ 0.61      &       89.92$\pm$ 0.35        \\
			STM+TSED+PT+2L &     50  &  80.09$\pm$ 0.49      &     49.14$\pm$ 0.62      & \textbf{86.95$\pm$ 0.44} &     92.34$\pm$ 0.49      & \textbf{92.96$\pm$ 0.62} &     82.73$\pm$ 0.25      &       89.83$\pm$ 0.30
	\end{tabular}}
	\caption{Results from our embedding distillation models on the evaluation sets in Table~\ref{tbl:data} using the CNN-based (the rows 5-10) and the LSTM-based teacher models (rows 13-18) along with the teacher models (*-400), baseline model (*-50) and previous distillation model (ENC).
		All models are tuned on the development sets and the best performing models are tested on the evaluation sets.
		Since neural models produce different results at any training due to the random initialization, five models are developed for each approach to avoid (un)lucky peaks, except for *-400 and *-50 where the results are achieved by selecting the best models among ten trials on the development sets.
		Each score is based on these five trials and represented as a pair of [Average $\pm$ Standard Deviation].}
	\label{tbl:comparison}
\end{table*}

\subsection{Network Configuration}
\label{ssec:hyperparameters}
Two types of teacher models are developed using Convolutional Neural Networks (CNN) and Long Short-Term Memory Networks (LSTM); comparing different types of teacher models provides more generalized insights for our distillation framework.
All teacher models use 400 dimensional word embeddings, and all student models are based on CNN.
The CNN-based teacher and student models share the followings: filter sizes = [2, 3, 4, 5], \# of filters = 32, dimension of the hidden layer right below the softmax layer = 50.
Teacher models add a dropout of 0.8 to the hidden layer, wheres student models add dropouts of 0.1 to both the hidden layer and the the projection layer.
On the other hand, the LSTM-based teacher models use two bidirectional LSTM layers. 
Only the last output vector is fed into the 50 dimensional hidden layer, which becomes the input to the softmax layer.
A dropout of 0.2 is applied to all hidden layers, both in and out of the LSTM.
For both CNN and LSTM ensembles, all 10 teachers share the same model structures with different initializations.
Although this limited teacher diversity, our ensemble method produces remarkably good results (Section~\ref{sssec:distillensemble}).

Each student model may consist of one or two projection layers.
The one-layered projection adds one 50 dimensional hidden layer above the embedding layer that transfers core knowledge from the original embeddings to the distilled embeddings.
The two-layered projection comprises two layers; the size of the lower layer is the same as the size of teacher's embedding layer, 400, and the size of the upper layer is 50, which is the dimension of the distilled embeddings.
This two-layered projection is empirically found to be more robust than various combinations of network architectures including wider and deeper layers from our experiments.\footnote{Among convolutional, relational, and dense-networks with different configurations, the dense-network with the reported configuration produces the best results.}

\subsection{Pre-trained Weights}
An autoencoder comprising a 50-dimensional encoder and a 400-dimensional decoder is used to pre-train weights for the two-layered projection in student models, where the encoder and the decoder have the same and inversed shapes as the upper and lower layers of the projection. 
Note that results by using pre-trained weights for the one-layered projection are not reported in Table~\ref{tbl:comparison} due to the limited space, but we consistently see robust improvement using pre-trained weights for the projection layers.

\begin{figure}[hbt!]
	\centering
	\includegraphics[width=\columnwidth]{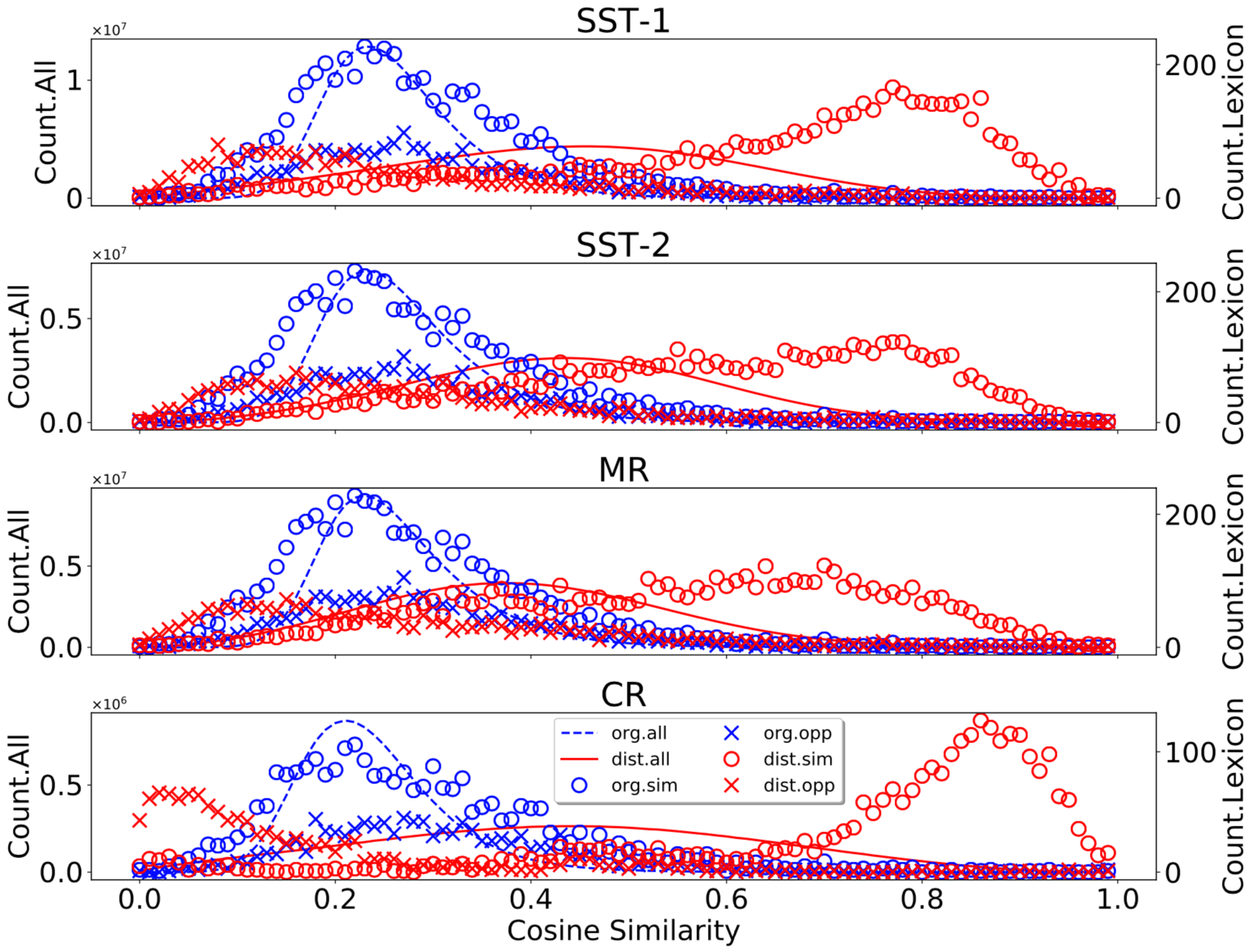}
	\caption{Similarity distributions among sentiment word pairs. 
		Blue and Red colors distinguish histograms from the original and distilled embeddings, respectively.
		The solid and dashed lines show the distributions from all word pairs regardless of their sentiments.
		Circles are for sentimentally similar word pairs, that are similarities between positive word pairs and negative word pairs such that $(w_i, w_j) \in (P_\ell \times P_\ell) | (N_\ell \times N_\ell)$. Crosses are for sentimentally opposite word pairs, that are similarities across positive and negative word pairs such that $(w_i, w_j) \in P_\ell \times N_\ell$.
	}
	\label{fig:lexicon}
	\vspace{-1ex}
\end{figure}


\subsection{Embedding Distillation}

Six models are evaluated on the seven datasets in Table~\ref{tbl:data} to show the effectiveness of our embedding distillation framework:
logit matching (LM), noisy logit matching (NLM), softmax tau matching (STM) models with teacher-student based embedding distillation (TSED), and another three models with the autoencoder pre-trained weights (*+PT) and the two layered projection network (*+2L).
Teacher models using 400-dim embeddings (*-400) are also presented along with the baseline model using 50-dim word embeddings (*-50) and the previous distillation model (ENC).
The comparison to these two models highlights the strength of our distilled models, significantly outperforming them with the same dimensional word embeddings (50-dim).

Table~\ref{tbl:comparison} shows the results achieved by all models.
While the two existing models, *-50 and ENC, show marginal differences, our proposed models, *+TSED, outperform the previous embedding distillation SOTA (ENC).
Our final models, *+PT+2L, outperform all the other models, reaching similar (or even higher) accuracy to the teacher models (*-400), which confirms that the proposed embedding distillation framework can successfully transfer the core knowledge from the original embeddings with respect to the target tasks, independent from the network structures of the teacher models.


The fact that the best model for each task comes from a different teacher-student strategy appears to be random.
However, considering that it is the nature of neural network models whose accuracy deviates at every training, this is not surprising although it signifies the need of ensemble approaches to take advantage of multiple teachers and produce a more robust student model.

\begin{figure*}[htbp]
	\centering
	\begin{subfigure}[b]{0.14\textwidth}
		\centering
		\includegraphics[width=0.99\textwidth]{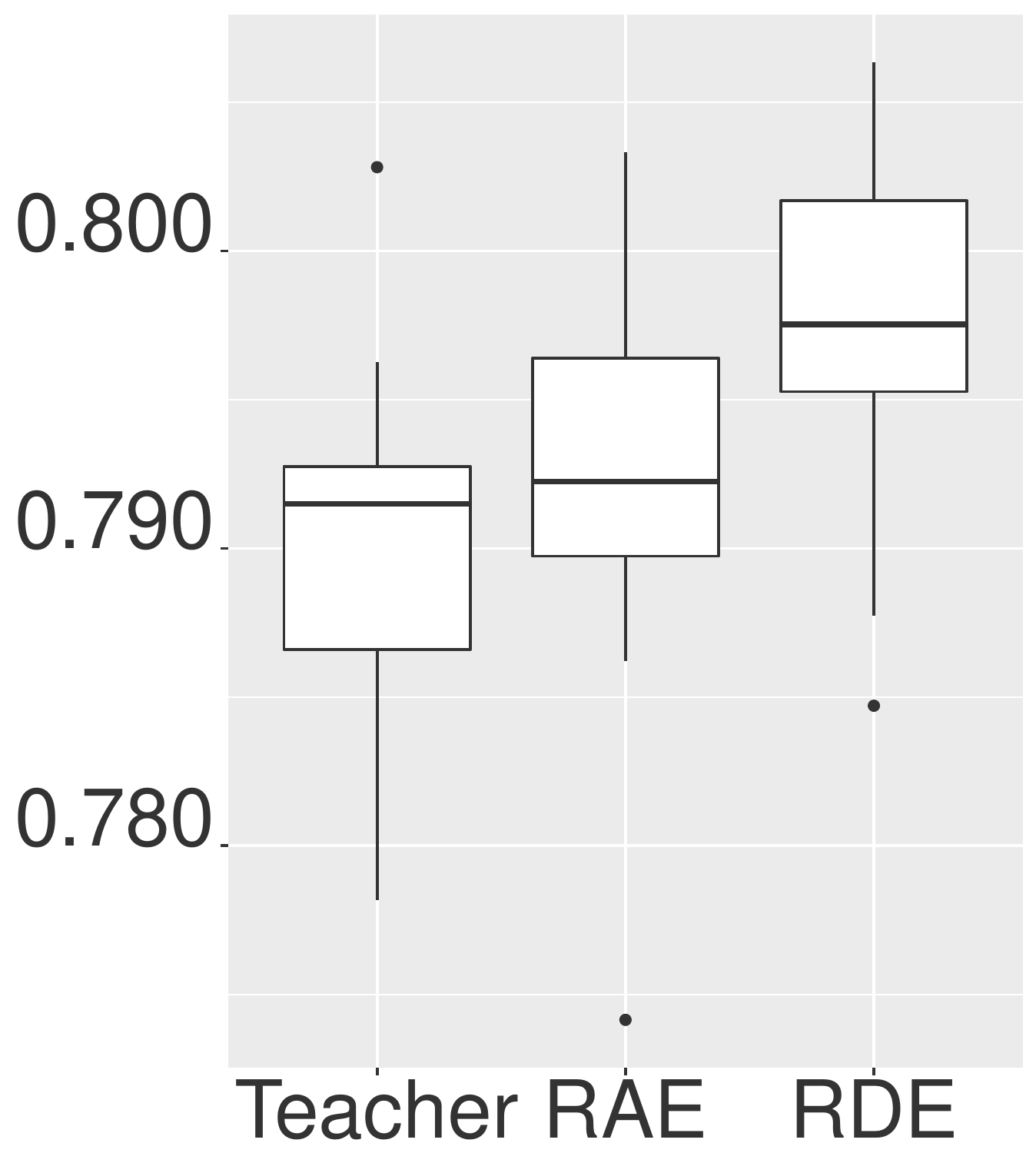}
		\caption{MR}
		\label{fig:ensemble.mr}
	\end{subfigure}%
	\begin{subfigure}[b]{0.14\textwidth}
		\centering
		\includegraphics[width=0.99\textwidth]{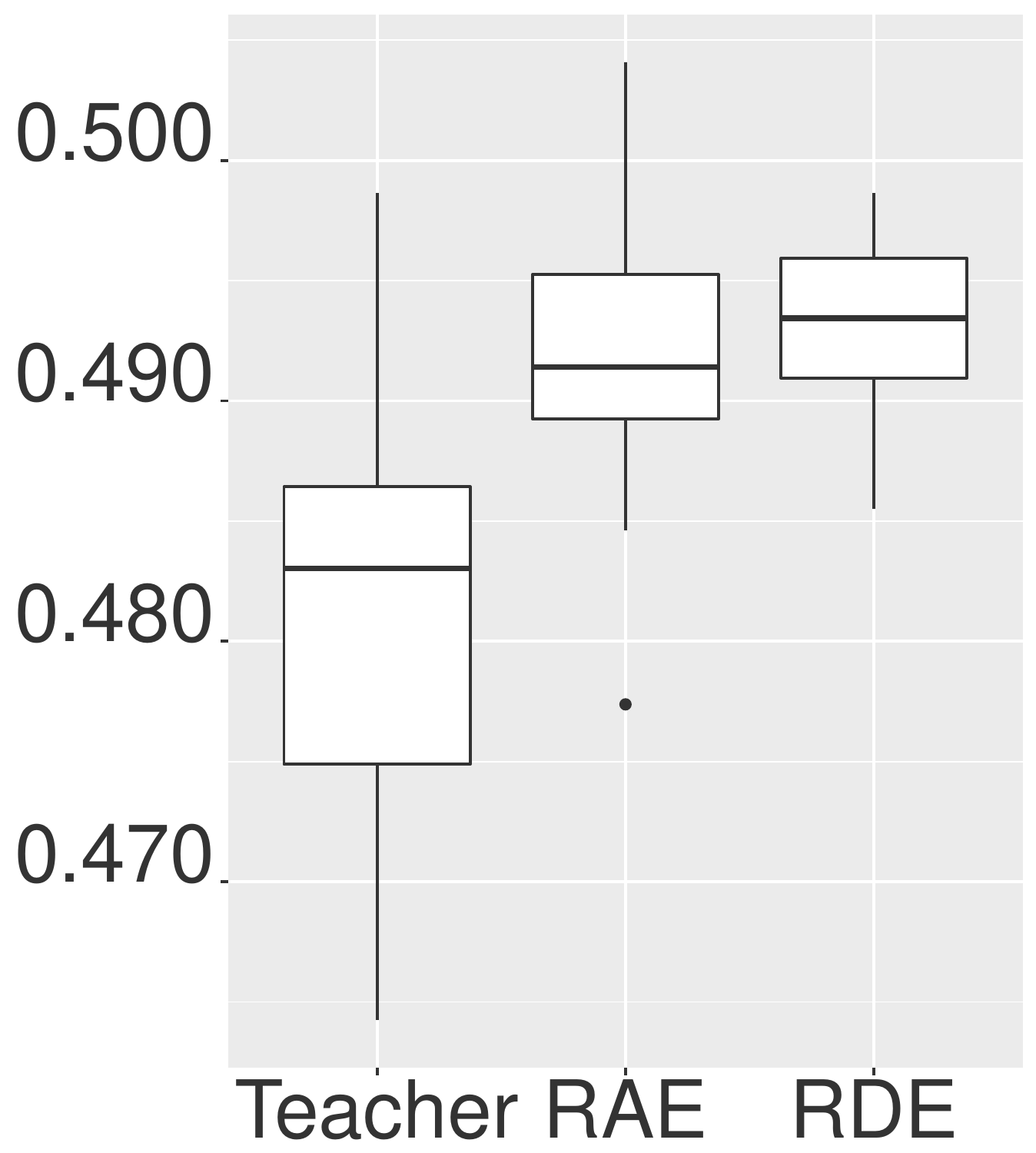}
		\caption{SST-1}
		\label{fig:ensemble.sst1}
	\end{subfigure}%
	\begin{subfigure}[b]{0.14\textwidth}
		\centering
		\includegraphics[width=0.99\textwidth]{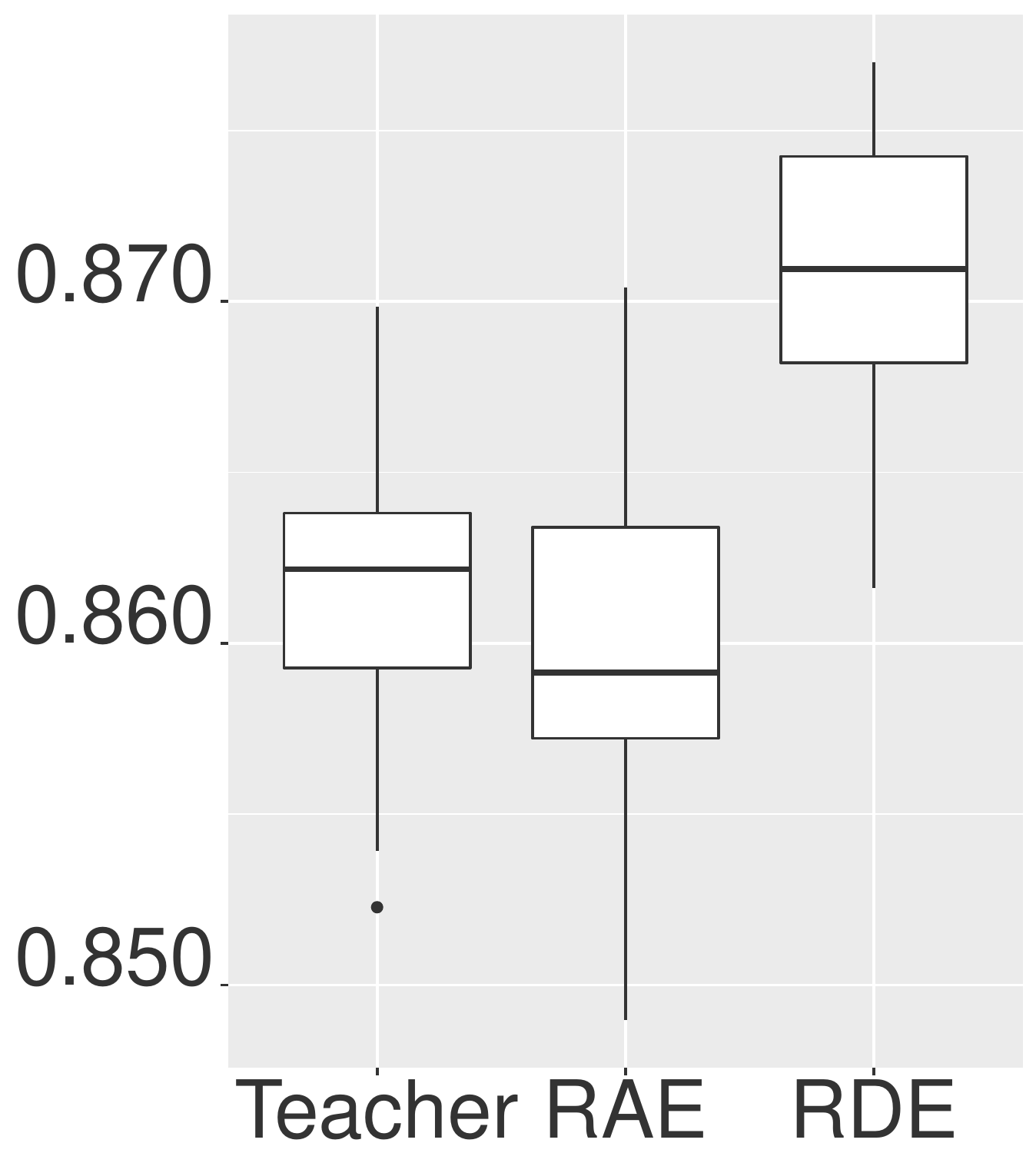}
		\caption{SST-2}
		\label{fig:ensemble.sst2}
	\end{subfigure}%
	\begin{subfigure}[b]{0.14\textwidth}
		\centering
		\includegraphics[width=0.99\textwidth]{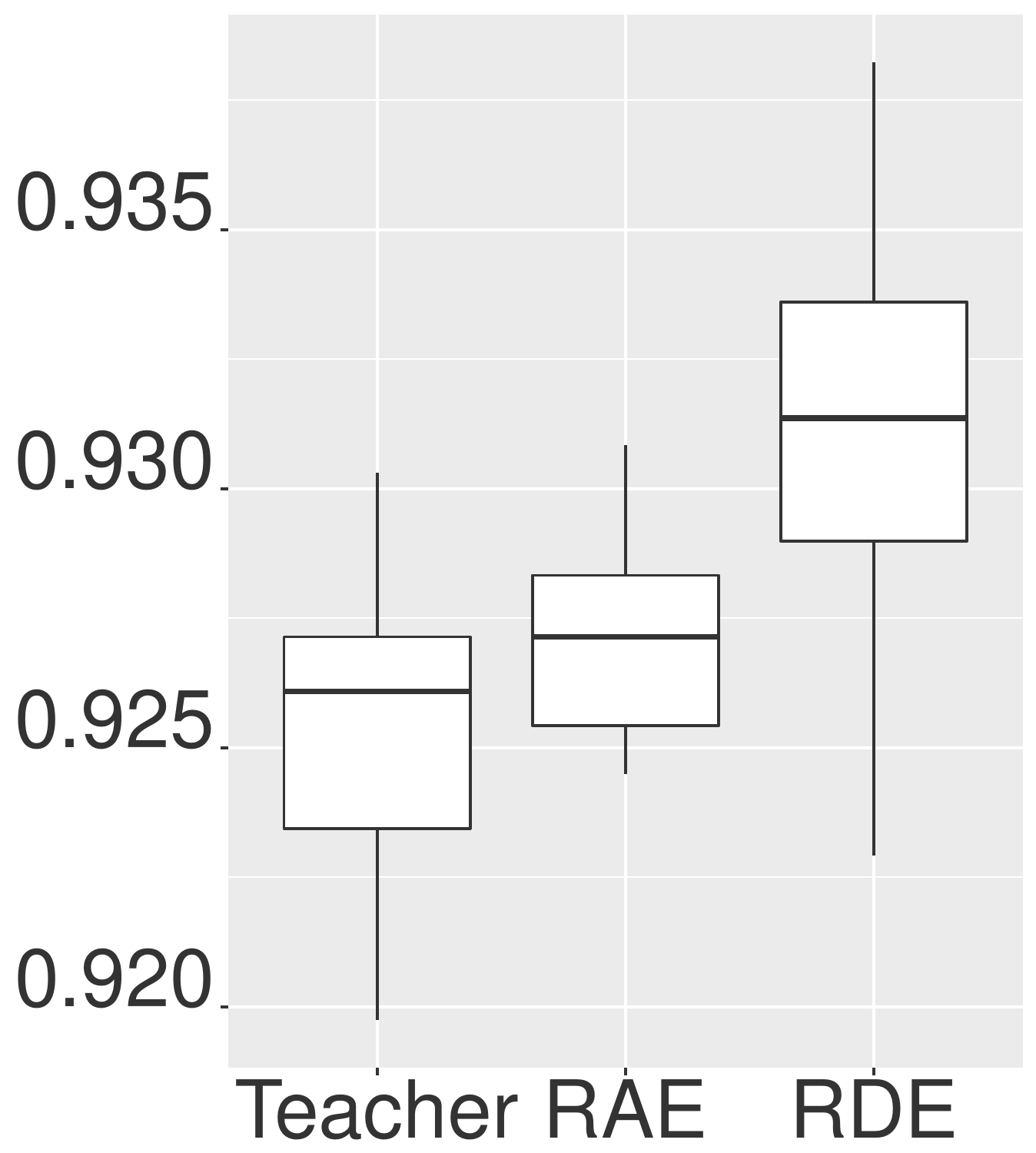}
		\caption{Subj}
		\label{fig:ensemble.subj}
	\end{subfigure}%
	\begin{subfigure}[b]{0.14\textwidth}
		\centering
		\includegraphics[width=0.99\textwidth]{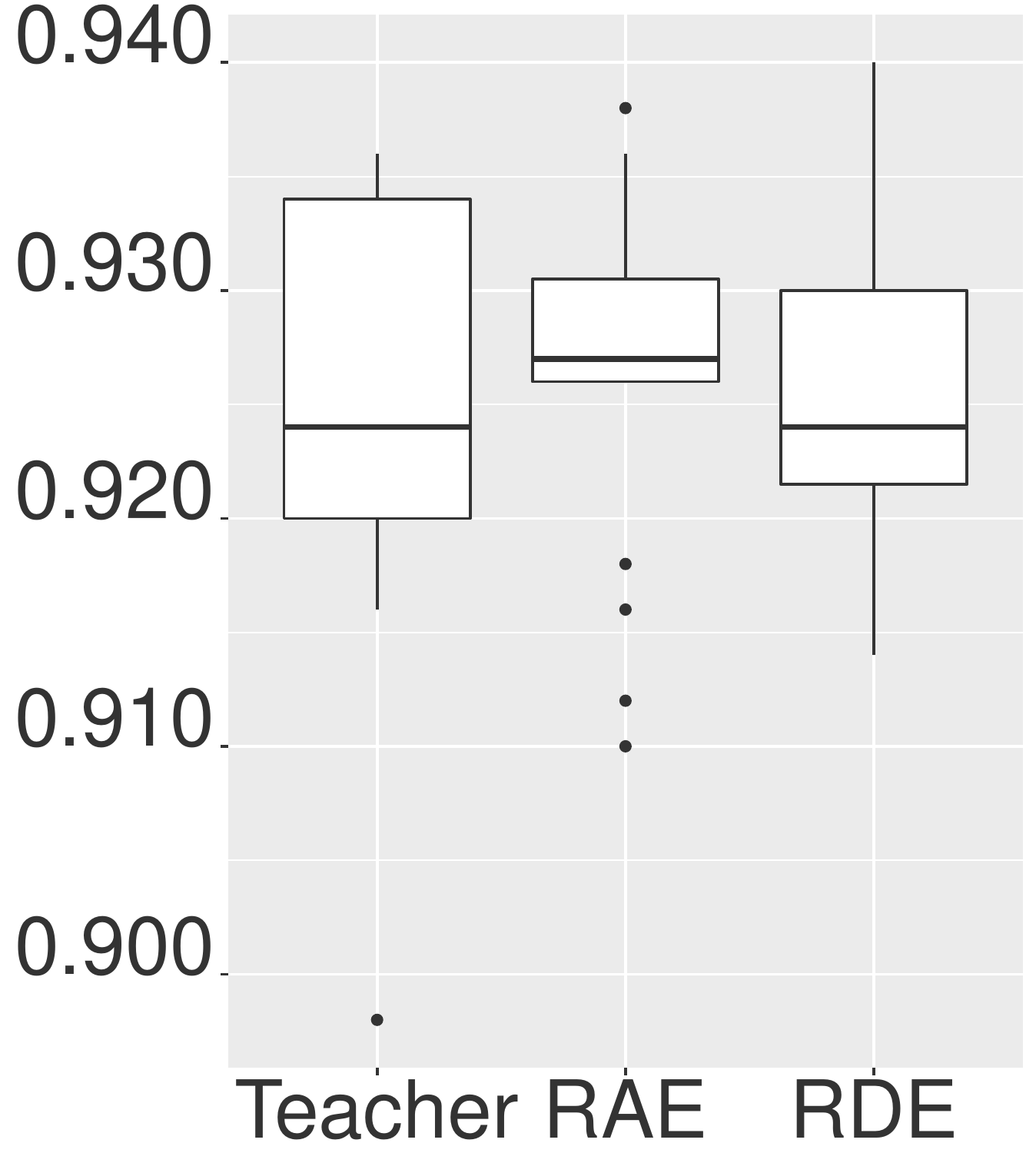}
		\caption{TREC}
		\label{fig:ensemble.trec}
	\end{subfigure}%
	\begin{subfigure}[b]{0.14\textwidth}
		\centering
		\includegraphics[width=0.99\textwidth]{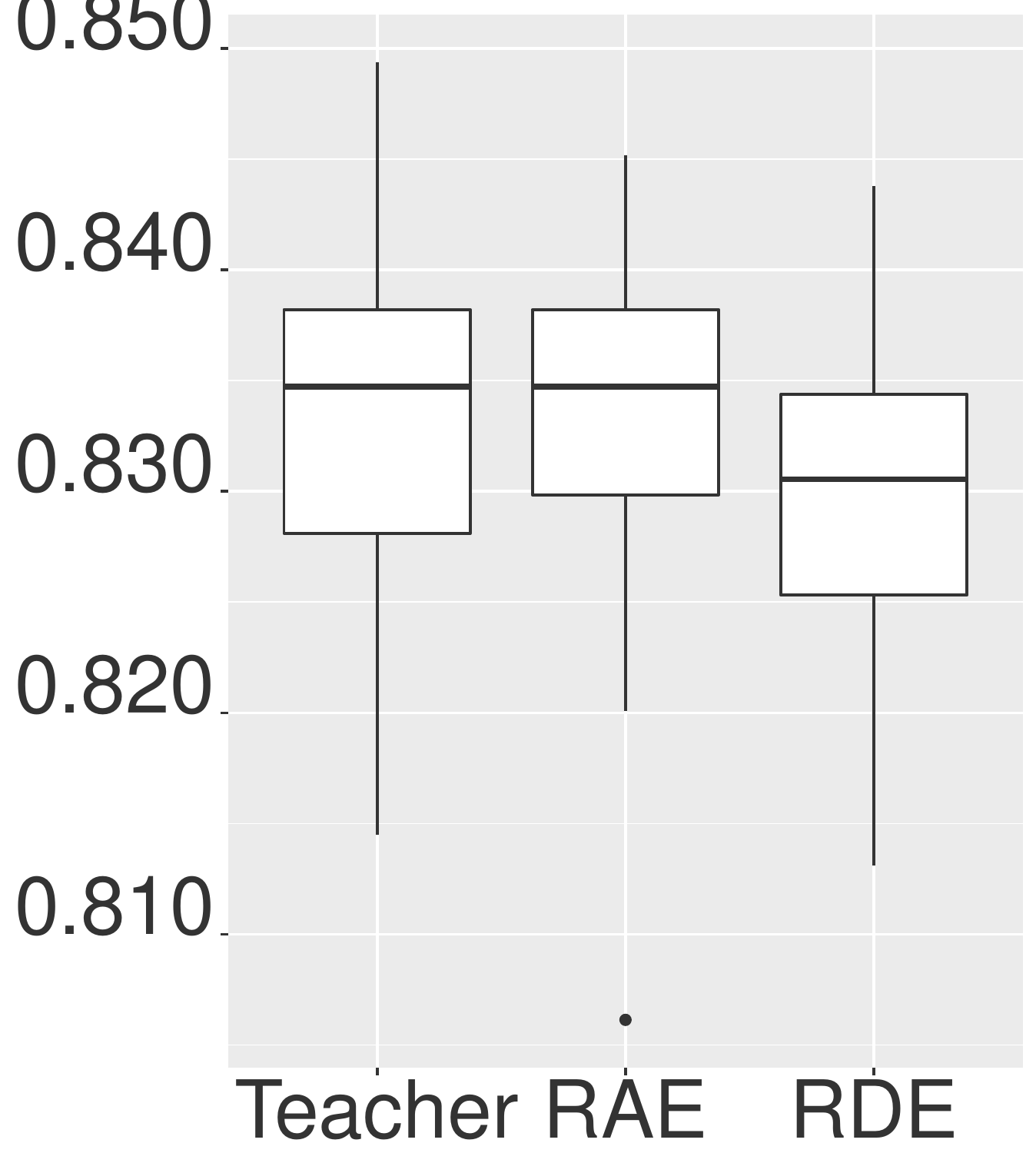}
		\caption{CR}
		\label{fig:ensemble.cr}
	\end{subfigure}%
	\begin{subfigure}[b]{0.14\textwidth}
		\centering
		\includegraphics[width=0.99\textwidth]{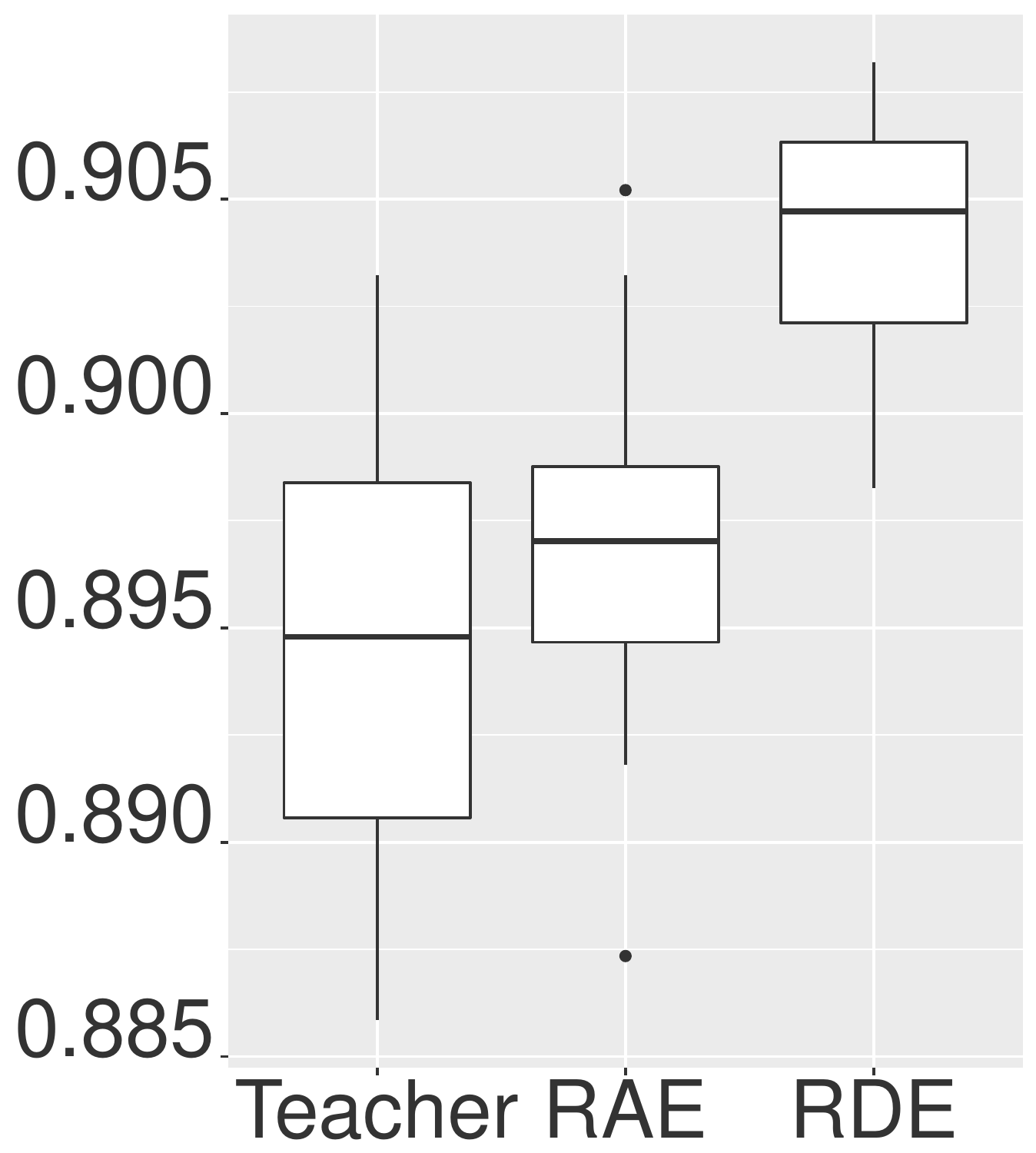}
		\caption{MPQA}
		\label{fig:ensemble.mpqa}
	\end{subfigure}%
	\caption{Accuracy comparisons between the ensemble and the teacher models.
		To avoid (un)lucky peaks, each method is evaluated 20 times where each trial produces a different result.
		These evaluation results are shown as boxplots in this figure.}
	\label{fig:ensemble}
	\vspace{-2ex}
\end{figure*}

\subsection{Lexical Analysis}
\label{sssec:lexiconanalysis}

Embedding distillation for a task such as sentiment analysis can be viewed as a vector transformation that adjusts similarities between word embeddings with respect to their sentiments.
Thus, we hypothesize that distilled word embeddings from sentiment analysis should bring similar sentiment words together while disperse opposite ones in vector space.

To verify this, lexical analysis on both the original and the distilled word embeddings is conducted using the four sentiment datasets: SST-1, SST-2, MR, and CR.
First, positive and negative words are collected from two publicly available lexicons, the MaxDiff Twitter Sentiment Lexicon~\cite{kiritchenko2014sentiment} and the Bing Liu Opinion Lexicon~\cite{hu2004mining}.
Then, two groups of sentiment word sets are constructed, ($P_t$, $N_t$) and ($P_o$, $N_o$), where $P_*$ and $N_*$ compose positive and negative words, and $*_t$ and $*_o$ are collected from the Twitter and the Opinion lexicons, respectively.
Next, the intersection between each type of sentiment word sets is found, that are $P_{to} = P_t \cap P_o$ and $N_{to} = N_t \cap N_o$, where $|P_{to}| = 72$ and $|N_{to}| = 89$.
Finally, the intersections between $*_{to}$ and the vocabulary set from each of the four sentiment datasets are found (e.g., $P_{\ell} = A_{\ell} \cap P_{to}$, where $\ell$ is one of the four datasets and $A_{\ell}$ is the set of all words in $\ell$):

\begin{itemize}
	\item $\ell = \mathrm{SST}_1 \Rightarrow |P_{\ell}| = 66$ and $|N_{\ell}| = 83$.
    \item $\ell = \mathrm{SST}_2 \Rightarrow |P_{\ell}| = 67$ and $|N_{\ell}| = 83$.
    \item $\ell = \mathrm{MR} \Rightarrow |P_{\ell}| = 67$ and $|N_{\ell}| = 82$.
    \item $\ell = \mathrm{CR} \Rightarrow |P_{\ell}| = 19$ and $|N_{\ell}| = 57$.
\end{itemize}

\noindent For each $\ell$, cosine similarities are measured for all possible word pairs $(w_i, w_j) \in PN_\ell \times PN_\ell$ where $PN_\ell = P_{\ell} \cup N_{\ell}$, using the original and distilled embeddings.
Figure~\ref{fig:lexicon} illustrates the similarity distributions.
It is clear that similar sentiment words generally give high similarity scores with the distilled embeddings (the plots drawn by red circles), whereas opposite sentiment words give low similarity scores (the plots drawn by red crosses).
On the other hand, low similarity scores are shown for any case with the original embeddings (the plots drawn by blue circles and crosses).
The normal distributions derived from the distilled embeddings (the red lines) are more symmetric and spread than the ones from the original embeddings (the blue lines), implying that distilled embeddings are more stable for the target task.


\subsection{Distillation Ensemble}
\label{sssec:distillensemble}

All ensemble models are based on our distillation framework using logit matching (LM) where the teacher models compose of 10 CNN-based or 10 LSTM-based models.
Two ensemble methods are evaluated: Routing by Agreement (RAE) and Routing by Disagreement (RDE), and
Figure~\ref{fig:ensemble} shows comparisons between these ensemble models against the teacher models.

The most notable finding is that RDE significantly outperforms the teacher, if the dataset is big.
For example, RDE outperforms the teacher models on average, except for CR and MPQA, whose training set is relatively small (Table~\ref{tbl:data}).
The insight behind this trend might be that if there are many data samples, then the probability of exploring different knowledge from minor opinions could be increased, which would positively affect to the ensemble.



\subsection{Model Reduction}
\label{sssec:modelsize}

The deployment models from either distillation or ensemble are notably smaller than the teacher models.
Since word embeddings occupy a large portion of neural models in NLP, reducing the size of word embeddings through distillation decreases the size of the entire model roughly by the same ratio as its reduction; in our case, eight times ($\nicefrac{400}{50}$).
Furthermore, if the proposed distillation ensemble method is compared to other typical ensemble methods, 
this reduction ratio becomes even larger.
Table~\ref{tbl:numparam} shows the number of neurons required for previous ensemble methods and the proposed one.
When training, the proposed one comprises 10\% more parameters due to the distillation process.
However, when deploying, the reduction of neuron is about eighty times ($\nicefrac{4000}{50}$).
This is because the proposed framework doesn't require repetitive evaluation of teacher models when testing, while other ensemble methods require evaluation of all sub models (teachers).


\begin{table}[htp!]
 	\centering
 	\resizebox{\columnwidth}{!}{
	\begin{tabular}{l||l|l|l}
		& \multicolumn{1}{c|}{\bf Previous} & \multicolumn{1}{c}{\bf Proposed} & \multicolumn{1}{c}{\bf Reduction}\\  \hline \hline
		Train       & $O(400M*10)$ & $O(400M*11)$ & $\times 0.91$ \\ 
		Deploy      & $O(400M*10)$ & $O(50M)$  & $\times 80$\\
	\end{tabular}}
	\caption{The number of neurons in previous ensemble methods and the proposed distillation ensemble method for training and deploying. $M$ represents the basic unit of model size for the embedding dimension 1. This table assumes ensemble with 10  teachers.}
	\label{tbl:numparam}
	\vspace{-2ex}
\end{table}


\noindent It is worth mentioning that the deployment models produced by distillation ensemble not only outperform the teacher models in accuracy but also are significantly smaller such that they operate much faster and lighter than the teacher models upon deployment.
This is very welcoming for those who want to embed these models into low-resource platforms such as mobile environments.

%% file: tex/conclusion.tex
\vspace{-0.5ex}
\section{Conclusion}
This paper proposes a new embedding distillation framework based on several teacher-student methods.
Our experiments show that the proposed distillation models outperform the previous distillation model and give compatible accuracy to the teacher models, yet they are significantly smaller.
Lexical analysis on sentiments reveals the comprehensiveness of the distilled embeddings.
Moreover, a novel distillation ensemble approach is proposed, which shows huge advantage in both speed and accuracy over any teacher model.
Our distillation ensemble approach consistently shows more robust results when the size of training data is sufficiently large.

